\documentclass{llncs}

\usepackage{graphicx}
\newcommand{\BF}[1]{\mbox{$\mathbf {#1}$}}
\usepackage{fancyhdr}

\begin{document}

\title{Dynamic Load Balancing Strategies for Hierarchical \emph{p}-FEM Solvers}
\titlerunning{Load Balancing for the \emph{p}-FEM}
\author{Ralf-Peter Mundani\inst{1} \and Alexander D\"uster\inst{1} \and
Jovana Kne\v{z}evi\'{c}\inst{2} \and Andreas Niggl\inst{3} \and Ernst Rank\inst{1}}
\authorrunning{Ralf-Peter Mundani et al.}
\institute{TU M\"unchen, Chair for Computation in Engineering, 80333 M\"unchen, Germany
\and
University of Belgrade, Department of Mathematics, 11000 Belgrade, Serbia
\and
SOFiSTiK AG, 85764 Oberschleissheim, Germany}

\maketitle

\begin{abstract}
Equation systems resulting from a \emph{p}-version FEM discretisation typically require a
special treatment as iterative solvers are not very efficient here. Applying hierarchical
concepts based on a nested dissection approach allow for both the design of sophisticated
solvers as well as for advanced parallelisation strategies. To fully exploit the underlying
computing power of parallel systems, dynamic load balancing strategies become an essential
component.
\end{abstract}

\pagestyle{empty}
\thispagestyle{fancy}
\lhead{}
\chead{}
\rhead{}
\lfoot{\scriptsize This is a pre-print of an article published in Ropo~M., Westerholm~J., Dongarra~J.\ (eds) Recent Advances in Parallel Virtual Machine and Message Passing Interface. EuroPVM/MPI 2009. Lecture Notes in Computer Science, vol 5759. The final authenticated version is available online at: https://doi.org/10.1007/978-3-642-03770-2\_37}
\cfoot{}
\rfoot{}

\flushbottom

\section{Introduction}
Within computational engineering, structure dynamics are one main source for challenging
numerical computations. Here, the \emph{p}-version of finite element methods is a very
prominent technique allowing to increase accuracy without increasing the amount of elements,
too. Nevertheless, the resulting equation systems fail to be efficiently solved with
iterative methods such as CG or multigrid and, hence, need to be processed via expensive
direct solvers (Gauss and relatives) that also entail limited potential for the
parallelisation. Applying hierarchical methods based on the nested dissection concept opens
the door to sophisticated solvers but also to new problems concerning the scalability of
parallelisation strategies related to tree structures. Hence, an optimisation of both
run time and parallel efficiency arises the necessity of dynamic load balancing strategies
that are able to exploit the underlying hierarchy and, thus, to leverage parallelisation
on all levels ranging from multithreading to distributed computing.

In this paper, we will show the hierarchical organisation of the \emph{p}-version using
octrees and their advantages for the solution of equation systems and parallelisation.
Furthermore, we will show a dynamic load balancing strategy that allows to tackle the
scalability problem which is essential --- for instance --- within interactive computational
steering applications in order to achieve small run times and, thus, high update frequencies.
As this paper describes work in progress, we will mainly highlight concepts and their
benefits for parallelisation instead of concrete benchmark results which are subject to
future research.

\section{Structural Analysis of Thin-Walled Structures}
The development of accurate and efficient element formulations for thin-walled structures
has been in the focus of research in Computational Mechanics since the advent of the finite
element method. At a very early stage it seemed to be clear that an investigation of plate
or shell problems with tetrahedral or hexahedral elements is not feasible for practical
problems, as a sufficient accuracy could only be obtained by a prohibitively large amount
of degrees of freedom and computational effort. The major reasons for this observation are
the mapping requirements of isoparametric, low order elements. Accurate solutions can only
be obtained if the ratio aspect of an element is close to one, resulting in an enormous
amount of elements, even if only one or a few layers are used over the thickness of the
structure. A natural consequence of this observation was to use dimensionally reduced
models like Reissner-Mindlin plates or Naghdi shells, and to build element formulations
based on these theories. However, it turned out that pure displacement type elements for
these models lead to notorious numerical problems like locking, giving rise to the
development of numerous improvements, like mixed elements, for example.

The approach presented in this paper is different from concepts usually applied when low
order elements are chosen. The idea is to construct a hierarchic family of high order
elements for both thin and thick-walled structures to make it possible to control the model
error inherent in every plate or shell theory by simply increasing the polynomial degree of
the trial or Ansatz functions in thickness direction. The high order finite element
approach for three-dimensional thin and thick-walled structures is based on a hexahedral
element, applying hierarchic shape functions
\cite{Szabo:91,Duester:01.2,Szabo:04.1,Duester:07.2}. The present implementation not only
allows the polynomial degree to be varied for the three different local directions but also
a different degree to be chosen for each primary variable, reducing the numerical effort
significantly.

\vspace*{-0.5cm}
\begin{figure}[h]
  \centering
  \includegraphics[width=9cm]{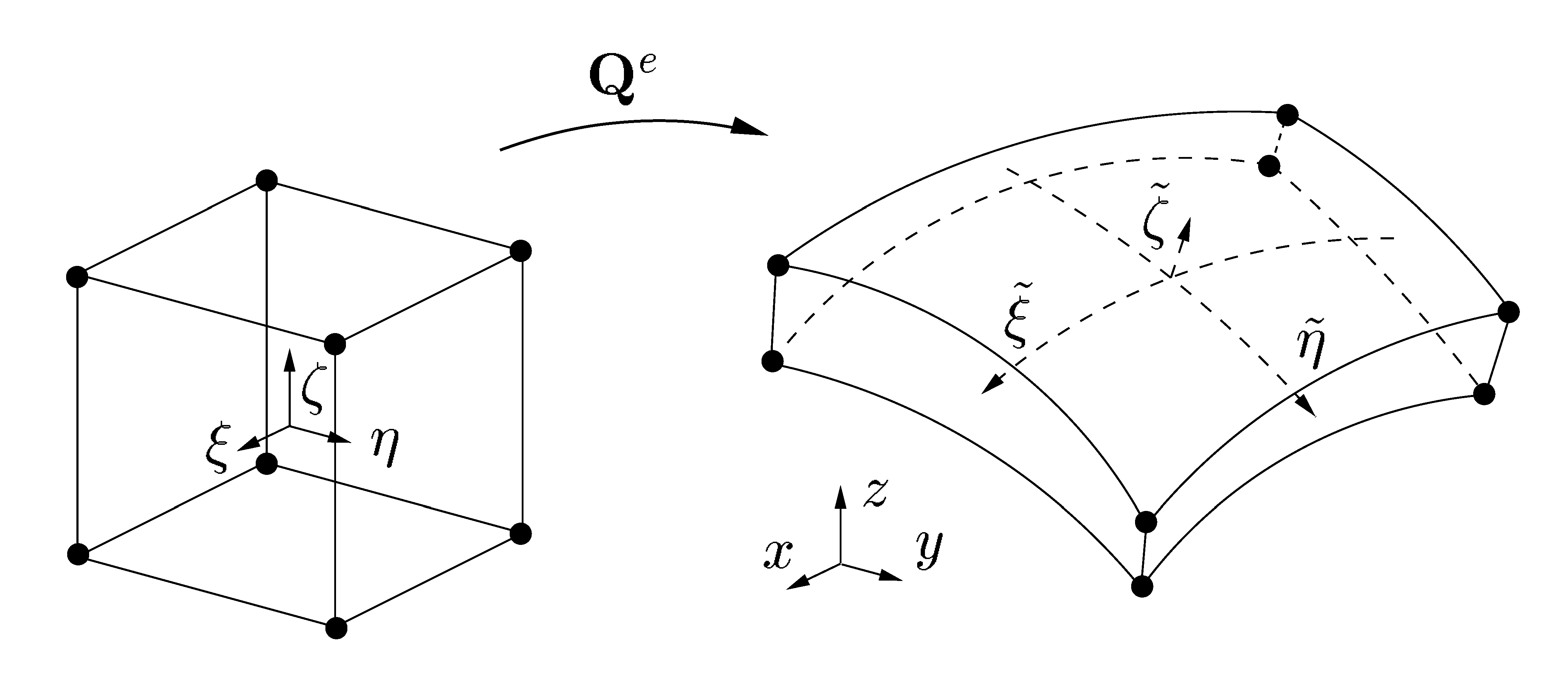}
  \caption{Curved shell-like solid element of high order}
  \label{fig:shell3d}
\end{figure}

\vspace*{-0.5cm}
Figure~\ref{fig:shell3d} depicts a hexahedral element, discretising a part of a thin-walled
structure. Since the blending function method is used (see
\cite{Szabo:91,Duester:01.2,Szabo:04.1}, e.\,g.) the geometry of the discretised domain may
be quite complex. The shell-like solid may, for instance, be doubly curved with a
non-constant thickness. When thin-walled structures like the one depicted in
Figure~\ref{fig:shell3d} are to be discretised, it is important to treat the in-plane
direction ($\xi,\eta$) and the thickness direction ($\zeta$) differently. This can be
accounted for by using anisotropic Ansatz functions for the three-dimensional displacement
field $\BF{u} = \left[u_x,u_y,u_z\right]^T$\,. In some situations it may be sufficient to
restrict the polynomial degree of the Ansatz in thickness direction ($\zeta$) to a certain
degree, for example, $q=3$ whereas the Ansatz chosen for the in-plane direction $(\xi,\eta)$
is to be of order $p \ge 3$\,. 

Since the \textit{p}-version is less prone to locking effects \cite{Szabo:91,Szabo:04.1}, a
pure, strictly three-dimensional displacement formulation can be applied. The numerical
effort related to the computation of thin as well as thick-walled structures based on this
formulation has two major sources: the computation of the element stiffness matrices and
the solution of the resulting linear equation  system. Comparing classical dimensionally
reduced low-order finite elements for plates or shells with the proposed high-order
formulation it turned out in many benchmark computations \cite{Duester:01.2,Szabo:04.1}
that the high-order approach needs much less degrees of freedoms for the same accuracy and
shifts the computational work from the global level (solution of equation system) to the
local level, i.\,e.\ the computation of element matrices which is due to the numerical
integration of the element matrices numerically quite demanding. Considering the
parallelisation of this approach this can be regarded as an important advantage since the
computation of the element matrices does not necessitate any communication and can, thus, be
parallelised very efficiently, see \cite{Rank:01.4}. Whereas the parallelisation of the
computation of the element stiffness matrices is straightforward, the solution of the
resulting linear equation system is more involved. This is due to the fact that the
parallel solution of the equation system can not be carried out without any communication
between the parallel processes. Furthermore, we are restricted in the choice of the solver
since we apply a strict three-dimensional element formulation which results in a equation
system with a poor condition number when discretising thin-walled structures. Therefore,
iterative procedures, such as the preconditioned conjugate gradient method, for example,
turn out to be not efficient. This drawback is even more pronounced when nonlinear problems
of structural mechanics (hyperelasticity, elastoplasticity, etc.) are considered, worsening
the condition of the equation system. We therefore prefer to apply a direct solver which
will be described in the next section.

\section{Hierarchical Organisation of the \emph{p}-Version}
As described in the previous section, iterative solvers are not very efficient due to the
poor condition number of the equation system. Even direct solvers are advantageous here,
they nevertheless suffer from a high computational complexity and they typically entail
extensive parallelisation strategies in order to exploit the underlying computing power.
Hence, different approaches are necessary to cover the aforementioned problems. Well-known
from the field of domain decomposition is the nested dissection method that was first introduced
by J.A.~George \cite{george}. The basic idea of nested dissection (ND) is to recursively subdivide
the computational domain and to set up a local equation system on each subdomain to be solved in a
bottom-up step by successively eliminating local influences (computing the so-called Schur
complement).

For our \emph{p}-version this means to start from the computed element stiffness matrices
and to organise those in a hierarchical way in order to apply ND. Therefore, octrees are used as
their underlying principle of spatial partitioning is advantageous for the hierarchical organisation
of the \emph{p}-version. Element stiffness matrices are stored to the octree's leaf nodes --- each
leaf node stores at most one element or stays empty --- while degrees of freedom (DOF) are stored to
leaf nodes and inner nodes. The corresponding node for storing a DOF can easily be determined by
finding the common parent node of all leaf nodes, i.\,e.\ element stiffness matrices, that share this
DOF. It's obvious that the closer some DOF is stored to the tree root node the later it will be
eliminated and the more processing time has to be invested \cite{rpm:ddm}. The position of a DOF in
the tree highly depends on the spatial partitioning of the domain. As octrees always halve each spatial
dimension in every step, structures with huge dimensions in length and small dimensions in width and
height, for instance, suffer from too many DOFs being concentrated in the root node. Here, a two-stage
approach will help that halves only in one spatial dimension as long as the size of the subdomain is
larger than some threshold value \cite{tru:ba08}.

The results achieved with ND so far are very promising, especially ND allows to vastly
reduce the computational complexity in case the underlying structure changes. As only those
tree branches that contain a modification (material parameters, e.\,g.) have to be
re-computed, all others stay untouched and the Schur complements that have been computed in
a previous run can be re-used \cite{rpm:diss}. Nevertheless, the computational effort for
complex scenarios is too high for retrieving results in real time --- as necessary within
interactive computational steering applications, for instance --- thus, parallelisation is
inevitable. Efficient strategies for the dynamic parallelisation of ND are subject of the
next section.

\section{Parallelisation Strategies}
For the parallelisation of our ND approach we want to address both multithreading and
distributed computing. This allows us to easily incorporate latest developments in
hardware such as multi- or manycore CPUs as well as to provide the necessary flexibility
towards a unique workload distribution which --- as we will see --- is quite difficult to
achieve and moreover plays a dominant role concerning fair speed-up and scalability values.
Therefore, starting from a classical tree parallelisation we will then advance to a
sophisticated dynamic load balancing strategy.

\subsection{Problem Analysis}
One main advantage due to the hierarchical organisation of ND via octress is the pure
vertical communication between parent and children nodes (upwards for sending Schur
complements and downwards for receiving the solution). Hence, cutting the tree at
some level $L$ --- defining $L = 0$ for the root level --- leads to $8^L$ independent
sub-trees which could be processed in parallel. Assuming now a full and balanced tree with
$N$ leaf nodes, i.\,e.\ $N$ element stiffness matrices, one would achieve the highest
parallelism for cutting this tree at level $L = \lceil\log_8 N\rceil - 1$\,. Nevertheless,
speed-up and efficiency are verly limited, as this approach is similar to the problem that
Minsky et al.\ posed in \cite{minsky} concerning the parallel summation of $2N$ numbers on
$N$ processors. As the amount of active processes decreases in our case by a factor of 8 in
each ND level, the possible values for speed-up and efficiency are slightly better,
nevertheless far away from being a satisfying result due to the huge amount of inactive
processes and the bad scalability inherent to this approach.

In order to achieve good results for both, i.\,e.\ speed-up and scalability, some efficient
load balancing strategy is inevitable. A master-slave approach will serve as starting
point here, nevertheless arising the necessesity for being adopted to the underlying problem.
First, when dealing with a large amount of slaves one single master might become the
bottleneck due to a huge communication advent, hence, a multi-level concept is required.
Furthermore, independent tasks have to be identified and according to their dependencies on
the results of other tasks then administrated by those masters. Tasks per se are processed
by the slaves in parallel and should incorporate (simultaneous) multithreading to speed-up
local computations. As this strategy also covers distributed computing, topics such as
distributed storage are of high relevance but not part of our work in the current stage.

\subsection{Task Management}
Before single tasks might be administrated by some master, an equal distribution of tasks
among all masters has to be initiated. Therefore, we choose a 2-level hierarchy with one
master on the first level and several masters on the second. To distinguish between those
masters, the one on the first level is called main master and the rest are called traders.
In case of more than two levels, only the masters on the lowest level are traders, the rest
are main master, second masters, third masters and so on. The difference between masters and
traders is that only the masters are serving requests from the slaves, delegating these
requests to the corresponding traders which then take care about the real data exchange.

To initiate a work load distribution, the main master starts to analyse the octree structure
and estimates the amount of work load, i.\,e.\ the amount of necessary elimination steps for
computing the Schur complement, in each node. Based on these values he is able to predict the
total amount of work and, thus, to decide how many traders and how many slaves should be
spawned. Furthermore, the estimated work load in each node allows the master to distribute
the octree among the traders more equally even in case of very imbalanced trees. This is not
the case when just cutting an imbalanced tree at some level $L$ and assigning the resulting
sub-trees to the traders. Nevertheless, to achieve an equal distribution the tree might be
cut into much more parts than traders, thus, one trader has to administrate several sub-trees
which do not always preserve neighbourhood relations. This might entail further complexity
due to more communication but could not be observed so far for the current implementation.

Once sub-trees have been assigned to a trader, the trader himself analyses the respective
tree structures in order to determine all tasks, i.\,e.\ the equation systems in all nodes,
together with their dependencies on child-nodes concerning the input data (Schur
complements). The tasks that have been identified together with their dependencies are then
stored to a priority queue which is updated by the trader each time some slave returns the
results of its computation. That means, tasks in the queue are checked if any of their
dependencies are fulfilled. Tasks without further dependencies are ready to be processed by
a slave and therefore get a higher priority while tasks that cannot be processed yet have a
priority $p = \infty$\,. The trader picks one task among those ready for processing and sends
the corresponding task ID to the master for being advertised to the slaves. The master stores
tuples consisting of task ID and trader ID --- one tuple per trader --- in order to serve
requests from the slaves. Hence, the master is not involved into heavy communications as he
just has to tell the slaves which trader to contact for which task.

\subsection{Processing Tasks}
Slaves always contact the master to request new tasks. If tasks are available, the master
picks one and sends the tuple task ID and trader ID as answer back to the slave before he
continues serving the next request. Receiving such a tuple, a slave can now contact the
trader and request the corresponding task for a local processing. Depending on the type of
task the trader initiates a data transfer, consisting of an element stiffness matrix or
several Schur complements. Again, within the current implementation we do not cover
distributed storage. In this case, the trader would send a mixture of locally stored data
and keys to remote repositories where the slave can retrieve the rest of the information.
When the data transfer has been finished, the slave holds all subsequent data to start its
computation without any further communication to the master or the trader.

If a slave receives an element stiffness matrix it can immediately start with the static
condensation of local DOFs in order to compute the Schur complement. In case it received
several Schur complements (as result from static condensation in the respective child-levels)
it first has to assemble its local equation system $K\cdot u = d$\,. Therefore, all Schur
complements $K_i$ are build together as $K = \sum_i K_i$ by summing up corresponding matrix
entries. Using several threads for the assembly step might entail heavy synchronisation in
case parallelisation takes places over the $K_i$ as no two threads are allowed to update the
same matrix element $k_{ij}$ in parallel. This can easily be solved deploying several critical
sections (one per column, row or some block of $K$) or by using all threads for processing
just one $K_i$ instead. While the latter one comes without the need for synchronisation it
entails lower parallelism due to the serial processing of all $K_i$\,. Concerning the static
condensation of local DOFs, a partial Gaussian elimination is performed. Here, a multithreaded
approach concerning the middle of the three nested loops from Gaussian elimination is advantageous,
as threads profit both from the independent loop iterations and the shared memory. Obviously,
this results in perfect, i.\,e.\ linear speed-up values on (SMP) architectures with 2, 4, and 8
cores as been tested. When finished, the slave contacts the same trader again to return its Schur
complement for the next level tasks.

In the current stage, slaves are implemented as memoryless processes, deleting all subsequent
information when the task has been finished. Keeping the regarding data even after the static
condensation saves communication time in the final solution step, as these parts do not have to be
transmitted again. Nevertheless, it implies a rigid order which slave has to process which task what
might lead to bottlenecks in some cases. 

\vspace*{-0.25cm}
\begin{figure}[h]
  \centering
  \includegraphics[width=6.0cm]{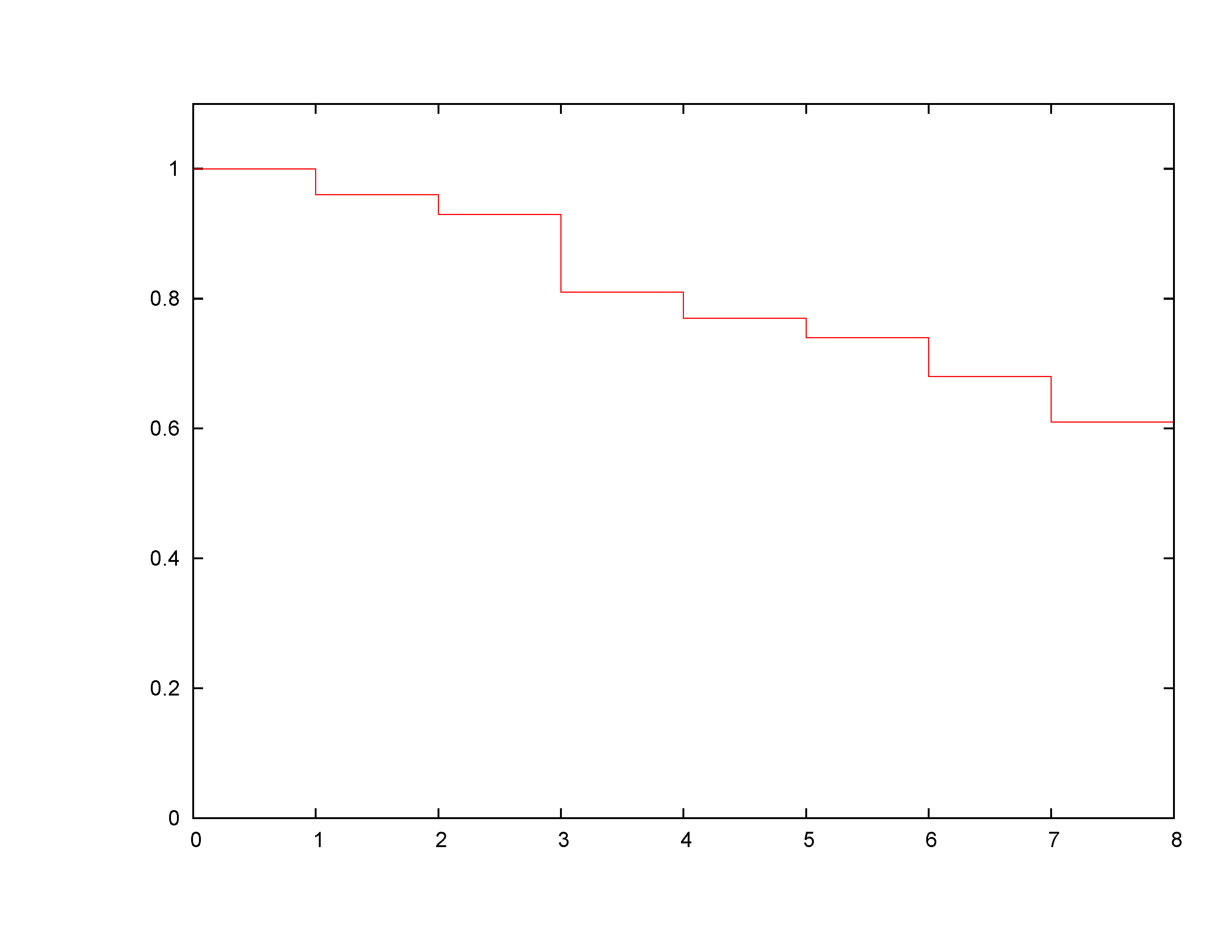}
  \includegraphics[width=6.0cm]{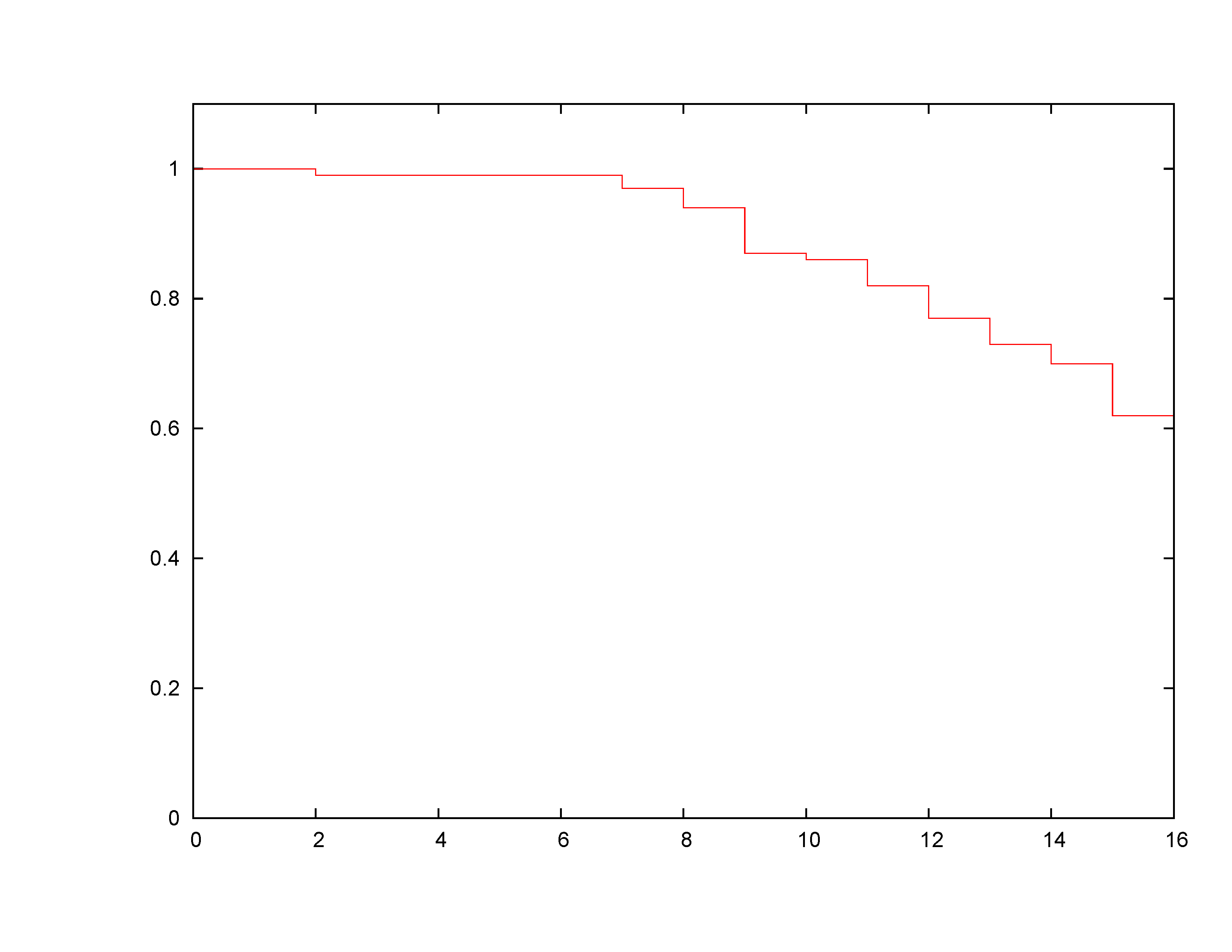}
  \caption{Normalised working index for 8 and 16 slaves processing a problem with 4171 initial
           independent tasks on leaf level}
  \label{fig:workidx}
\end{figure}

\vspace*{-0.5cm}
A much more interesting questions is related to the working index of slaves. Slaves are either
active, i.\,e.\ processing a task, or idle, i.\,e.\ waiting to be served. Summing up all phases a
slave \emph{i} is active and dividing it by the slave's total processing time results in a normalised
working index $\omega_i \in [0,1]$. Plotting all $\omega_i$ in decreasing order provides a step function
that allows to estimate about the mean time slaves are busy. In the ideal case, all $\omega_i$ are close
to 1, but due to the problem of decreasing activities (i.\,e.\ independent tasks) in each higher ND level,
this will be the rare case. Figure~\ref{fig:workidx} shows the working index for 8 and 16 slaves
processing a problem with 4171 elements, i.\,e.\ inital independent tasks, served by 4 and 8 traders,
resp. It can be observed, that the working indices vary between 0.6 and 1 which is a promising result
especially when compared to the theoretical values according to Minsky et al. Nevertheless, only about
half of the processes are active more than 90\% of the total processing time, which is somewhat clear as
the task size increases on each ND level when approaching the tree root and, thus, inactive processes need
to wait longer before being served a new task. Here, further optimisation is possible if tasks larger than
some threshold size are distributed among several processes in order to be processed in parallel on both
process and block (via threads) level. This approach is closely coupled with the question at which point
distributed parallel processing is more efficient than multhithreading---a question that cannot be
answered easily regarding latest trends in multi- and manycore architectures and which is still part of our
researches.

\section{Conclusion}
In this paper, we have proposed a dynamic load balancing strategy for linear equation
solvers based on the nested dissection approach in order to tackle known problems related
to tree parallelisations. Applied to the \emph{p}-version of finite element methods, first
results sound very promising, bringing us one step closer to the long-term objective of
structure dynamics in real time as needed for interactive computational steering scenarios.

\section{Acknowledgements}
Parts of this work have been carried out with the financial support of the International
Graduate School of Science and Engineering (IGSSE) at Technische Universit\"at M\"unchen.

\end{document}